\documentclass[journal=aaembp,manuscript=article]{achemso}

\setkeys{acs}{maxauthors=0}

\usepackage{graphicx}
\usepackage{bm}
\usepackage{subcaption}
\usepackage{xcolor}
\usepackage{amsmath}
\usepackage{mathtools}
\usepackage[normalem]{ulem}

\usepackage{xr-hyper}
\title{Mesoscopic Modeling of Structure-Transport Relationships in Dense CNT Films Containing Amorphous Carbon}

\author{Yvelin Giret}
\affiliation{Institut des Molécules et Matériaux du Mans, UMR CNRS 6283, Le Mans université, 72085 Le Mans, France}
\alsoaffiliation{Nanolayers Research Computing Ltd., 51 New Way Road, London, England, NW9 6PL}
\alsoaffiliation{Department of Physics and Astronomy, University College London, Gower Street, London WC1E 6BT, United Kingdom}

\author{Filippo Federici Canova}
\affiliation{Nanolayers Research Computing Ltd., 51 New Way Road, London, England, NW9 6PL}
\alsoaffiliation{Department of Applied Physics, Aalto University, 00076 Aalto, Espoo, Finland}

\author{Al-Moatasem El-Sayed}
\affiliation{Nanolayers Research Computing Ltd., 51 New Way Road, London, England, NW9 6PL}
\alsoaffiliation{Institute for Microelectronics, Technische Universität Wien, A-1040 Vienna, Austria}

\author{Thomas R. Durrant}
\affiliation{Nanolayers Research Computing Ltd., 51 New Way Road, London, England, NW9 6PL}
\alsoaffiliation{Department of Physics and Astronomy, University College London, Gower Street, London WC1E 6BT, United Kingdom}

\author{Rahul Sen}
\affiliation{Nantero Inc., 25-B Olympia Avenue,Woburn, MA 01801, USA}

\author{Harry Luan}
\affiliation{Nantero Inc., 25-B Olympia Avenue,Woburn, MA 01801, USA}

\author{Gennadi Bersuker}
\affiliation{M2D Solutions, 4620 Trail West Dr., Austin TX, 78735, USA}

\author{Alexander L. Shluger}
\email{a.shluger@ucl.ac.uk}
\affiliation{Department of Physics and Astronomy, University College London, Gower Street, London WC1E 6BT, United Kingdom}
\alsoaffiliation{London Centre for Nanotechnology, University College London, Gower Street, London WC1E 6BT, United Kingdom}
\alsoaffiliation{WPI-Advanced Institute for Materials Research (WPI-AIMR), Tohoku University, 2-1-1 Katahira, Aoba-ku, Sendai 980-8577, Japan}

\author{David Z. Gao}
\email{david@nanolayers.com}
\affiliation{Nanolayers Research Computing Ltd., 51 New Way Road, London, England, NW9 6PL}
\alsoaffiliation{Department of Physics, NTNU NO-7491 Trondheim, Norway}
\alsoaffiliation{Department of Physics and Astronomy, University College London, Gower Street, London WC1E 6BT, United Kingdom}


\begin{document}
\maketitle

{\small \textit{Please cite as : ACS Appl. Electron. Mater. 2026, 8, 4324-4335}}

%\newpage

\begin{abstract}
 Carbon nanotube (CNT) films  are widely considered as prospective building blocks for advanced electronic and nanostructured materials. In particular, electrical transport in high-density CNT films results from a complex interplay between network morphology and CNT connectivity, which remains challenging to characterize quantitatively. To identify the structural parameters that govern the electrical current in CNT films, we employed coarse-grained molecular dynamics to construct dense mesoscale CNT film models that include CNTs with different chiralities and lengths. The effects of CNT geometrical features on the film morphologies were quantified by devising a set of structural descriptors and analyzing their mutual correlations. The impact of varying the concentration of amorphous carbon (aC) inclusions on the film structure was assessed. Finally, we employed a nodal analysis framework to compute the electrical current across the networks and correlate the charge transport characteristics to the underlying structural descriptors. The current is found to be enhanced in films that exhibit high curvature and buckling, low bundling, and strong connectivity. We discuss how the presence of aC inclusions modifies these morphological and current characteristics. This work provides a mesoscale modeling framework for modeling  structure-transport relationships in dense CNT films and highlights the role of morphological descriptors in guiding the interpretation of electrical transport in complex nanostructured networks.
\end{abstract}

\section{Introduction}\label{sec:intro}

Carbon-based electronics is widely regarded as a promising alternative to conventional silicon-based devices~\cite{avouris2007carbon,Hu2010CNTrev,chae2014carbon,rao2018carbon,khan2020review,maruyama2022physics,bharadwaj2025transitioning,Zhang2025HPC}. In particular, carbon materials such as carbon nanotubes (CNTs) and graphene exhibit exceptional electronic, optoelectronic and mechanical properties~\cite{dinadayalane2010remarkable,wang2013carbon,kumar2021review}. The extremely high predicted tensile strength and Young's modulus of CNTs make them among the strongest and stiffest known materials~\cite{takakura2019strength}. Although the properties of single CNTs and small ensembles of CNTs have been extensively studied in recent decades, both experimentally and theoretically, the properties of CNT films, which are strongly affected by the formation of bundles and junctions~\cite{Hu2010CNTrev,evans2012inter,feng2018quantitative,yamaguchi2019one,wang2021phenomenological,kim2022nanoscale}, remain comparatively less understood~\cite{chen2025controlled,xiang2025electrical}. 

Experimentally, it has been shown that the properties of CNT films can strongly depend on their density~\cite{zhang2005spark,aliev2007thermal,yang2010tuning,pohls2012physical,zhang2012high,kumanek2019thermal}. Various strategies have been developed to increase the density of CNT films, such as annealing to increase ordering and crystallinity~\cite{zhang2005spark,yang2010tuning} or mechanical pressure to promote tighter packing~\cite{zhang2012high}. For instance, buckypapers (\textit{i.e.}, macroscopic aggregates of carbon nanotubes) exhibit almost linear enhancement of electrical and thermal conductivity under compression~\cite{zhang2012high}.
The fabrication method also plays an important role, as sequential deposition of CNT layers can be used to tune the properties of the final film~\cite{Rathanasamy2021,wang2013carbon}.
Finally, the presence of amorphous carbon between and around CNTs is inherent to the CNT growth process, yet its effect on the physical properties of the films remains poorly understood.

Recently, CNT films have emerged as promising candidates for resistive random access memory (RRAM) and neuromorphic applications.~\cite{rueckes2000carbon,gilmer2018nram,YANG2024neuromorphic}. In these applications, CNT films are used as a conductive medium, however, our understanding of the mechanisms of electron transport through CNT films remains incomplete due to their intrinsically disordered structure, heterogeneous connectivity, and presence of amorphous carbon.

Mesoscale modeling offers an attractive framework to address these issues~\cite{zhigilei2005mesoscopic,volkov2010scaling,Volkov2010a,volkov2012heat}. It allows the direct incorporation of structural complexity, such as bundling, curvature, and intertube junctions into current-carrying networks, without relying on computationally expensive atomistic simulations or overly simplified continuum models. In particular, the use of structural descriptors may provide a compact and interpretable representation of key morphological features, thereby enabling a systematic exploration of how geometry and topology influence electron transport.

Early theoretical work has highlighted the importance of structural morphology in determining the physical properties of CNT films~\cite{kumar2005percolating,behnam2007computational}. However, these studies focused on two-dimensional (2D) networks and did not include an atomistic description of the film structure. More recently, computational modeling has been used to understand the behavior of CNT films at a mesoscopic level. The methodology can generally be divided into two classes: (\textit{i}) the use of a mesoscopic tubular (MT) potential that describes inter-CNT interactions based on a continuous interatomic potential~\cite{zhigilei2005mesoscopic,volkov2010structural,Volkov2010a,volkov2012effect}, and (\textit{ii}) the mesoscopic distinct element method (MDEM), which enables simulation of interactions between particles of arbitrary shapes~\cite{Ostanin2013,Wang2017,ostanin2018toward,Wang2018}. Simulations using these mesoscopic models have shown that various factors, particularly those related to the underlying structure of the network, strongly affect its mechanical and thermal properties~\cite{volkov2010structural,Lee2009,volkov2012heat,volkov2012effect,Wang2018}. In contrast to the 2D approximation representing CNTs as straight lines, the MT potential allows one to directly study the effects of CNT bending and bundling, as well as construct more complex structures~\cite{volkov2010scaling,Volkov2010a,volkov2010structural,Wittmaack2018,wittmaack2018mesoscopic,wittmaack2019phase}. For example, it has been shown that increased bending energy leads to an overall increase in film stability, defined as the tendency not to disintegrate into cellular structures consisting of individual bundles weakly connected to each other~\cite{volkov2010structural}. 

In this study, we focus on creating and analyzing high-density films with different CNT chirality, length, density, and amorphous carbon (aC) content to investigate whether such films can be rationally designed to achieve specific target morphologies. Unlike previous studies that focused on relatively low-density systems, we consider films with density approaching the close-packed limit and study both single-layer and stacked multilayer structures, similar to those used in experimental studies \cite{gilmer2018nram,sen2020engineering}. We employ a mesoscale modeling approach capable of capturing mechanical deformation and local disorder. By quantifying each structure through a set of physically motivated descriptors, we establish correlations between structural properties and electron transport efficiency. Moreover, we develop a simple approach to mimic the structural effect of amorphous carbon particles inside the films. This work sheds light on the structural factors that govern conductivity in complex CNT networks and provides a foundation for future modeling of electrical transport phenomena in dense CNT films and related nanostructured networks.

The paper is organized as follows. We discuss the choice of structural parameters, such as tube length and film density, in Section~\ref{method_structural_parameters}. The mesoscopic model and the incorporation of aC into the simulations are presented in Section~\ref{method_mesoscopic_model}. The procedure for constructing the films is presented in Section~\ref{method_CNT_films}. We define the set of descriptors in Section~\ref{method_metrics}. The nodal analysis used to compute the electrical current through the films is detailed in Section~\ref{method_current}. The results of the simulations are presented and discussed in Section~\ref{sec:results} followed by discussion in section~\ref{sec:Discussion} and concluding remarks in Section~\ref{sec:conclusion}. Additional data and supporting figures are provided in the Supporting Information (SI).

\section{Methodology}\label{sec:methods}
\subsection{Structural parameters}\label{method_structural_parameters}
Modeling of 3D CNT assemblies is challenging due to the huge variety of CNT types (single- and multiwall), chirality, lengths, and the complexity of their interaction and the morphology of the films, which includes CNT bending and bundling \cite{Hahm2012Bundling}. Driven by particular applications, it is inevitable to consider specific systems, characterized by a particular film density, morphology, and distribution of CNT lengths and chiralities. In particular, the film morphology is directly determined by the method of deposition and treatment. Here, we are guided by CNT films produced for RRAM devices fabricated as described in Refs~\cite{gilmer2018nram,sen2020engineering}. However, the methodology and analysis will be useful for future simulations of dense CNT films for other nanoelectronic applications.

Due to the specifics of the growth and deposition techniques, the CNT films considered in our work consist of stacked layers with different CNT lengths and chirality distributions~(see SI). Based on the experimentally measured distributions, we selected three representative lengths: short (15~nm),  intermediate (40~nm), and long (100~nm, corresponding to half the size of the simulation box), as well as two different chiralities, (16,0) and (32,0), each associated with a specific diameter: $d^{(16)} \simeq1.25~\mathrm{nm}$ and $d^{(32)} \simeq2.5~\mathrm{nm}$. These chiralities were chosen to represent the average diameter of CNTs observed in RRAM devices (see Fig.~S2) independently of the metallic or semiconducting character of the CNTs. As discussed below, the nodal analysis used to compute the electrical current depends on CNT diameter only, and not on the specific chiral indices.

Experimentally fabricated films are often composed of multiwalled CNTs, although RRAM devices have also been successfully realized using single-walled CNTs, indicating that multiwalled structures are not a strict requirement for device operation. In the present study, only single-walled CNTs are considered, as interactions between multiwalled CNTs are not currently implemented in the coarse-grained model. We chose a high density of 0.6~g$\cdot$cm$^{-3}$ (the close-packed density for (32,0) tubes is $\sim$0.8~g$\cdot$cm$^{-3}$), and also simulated films at the lower density of 0.3~g$\cdot$cm$^{-3}$ for comparison. The films were constructed in single-layer or four-layer stacked configurations. Finally, the amorphous carbon content was experimentally estimated to be between 10 and 40\% by mass. Based on two reference configurations, namely 100~nm CNT length, single-layer, 0.6~g$\cdot$cm$^{-3}$, and either (16,0) or (32,0) chiralities, the amount of aC was varied from 0\% to 40\% with a 10\% increment. By keeping the target density constant, the number of CNTs in the simulation box decreases as the amount of aC particles increases. The standard deviation of the final density is 7$\times$10$^{-4}$~g$\cdot$cm$^{-3}$ for the 0.3~g$\cdot$cm$^{-3}$ samples and 4$\times$10$^{-4}$~g$\cdot$cm$^{-3}$ for the 0.6~g$\cdot$cm$^{-3}$ samples. In total, 32 distinct structures were constructed and analyzed.

\subsection{Mesoscopic Model}\label{method_mesoscopic_model}

CNTs are modeled as collections of point particles (beads), each representing a 1.0~nm-long cylindrical segment of a CNT; the terminal beads are lighter and only represent a 0.5-nm~segment. Beads within the same CNT are connected by harmonic stretching and bending potentials, while beads from different, randomly oriented CNTs interact through nonbonded Lennard-Jones potentials, according to the mesoscopic tubular force field  developed by Volkov, Zhigilei, and collaborators~\cite{zhigilei2005mesoscopic,volkov2010structural,Volkov2010a,volkov2010scaling}. All calculations were performed with the \texttt{MESONT} package~\cite{kateris2020collisions,kloza2024carbon} as implemented in the \texttt{LAMMPS} code~\cite{LAMMPS}.

\begin{figure*}[t!]
\centering
\begin{minipage}[t!]{0.48\textwidth}
\centering
\includegraphics[width=\linewidth]{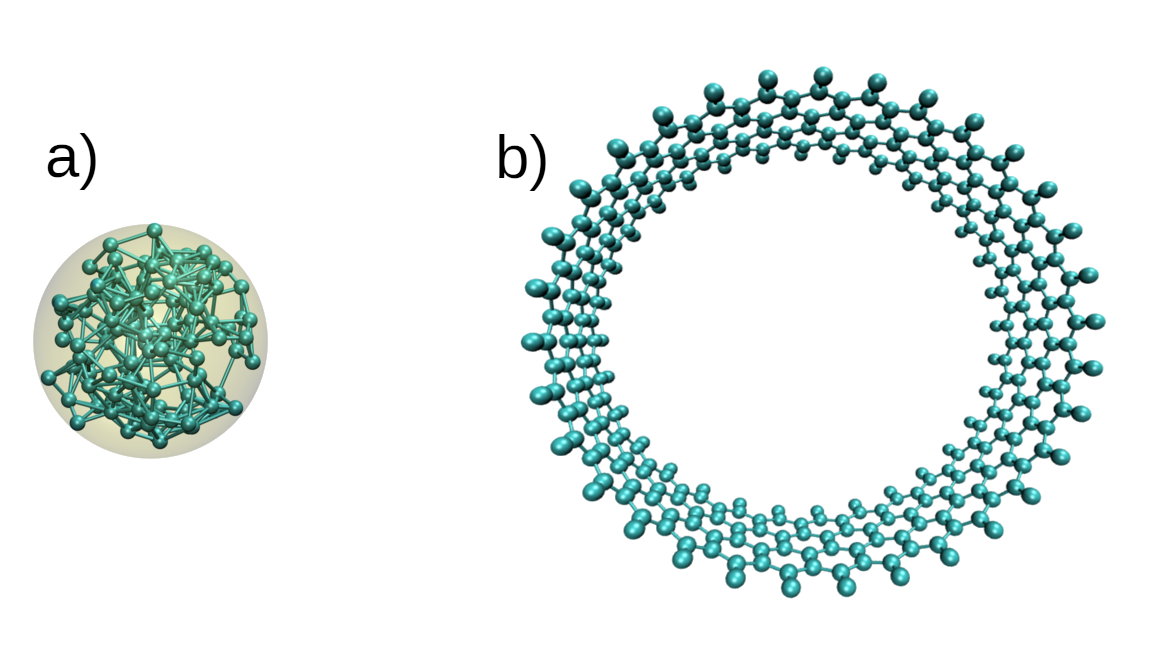}
\caption{Atomistic models of aC (a) and CNT (32,0) segment (b) used to compute the Lennard-Jones parameters describing their interactions.}
\label{fig:aCmodel}
\end{minipage}%
\hfill
\begin{minipage}[t!]{0.48\textwidth}
\centering
\begin{tabular}{l | c c}
        \hline\hline
        \multicolumn{1}{l}{\bfseries entity pair} & \multicolumn{1}{c}{\bfseries $\epsilon$ [eV]} & \multicolumn{1}{c}{\bfseries $\sigma$ [nm]} \\ \hline\hline
        aC--aC & 0.180 & 1.327 \\ \hline
        aC--CNT(16,0) & 0.386 & 1.420 \\ \hline
        aC--CNT(16,0)$\dag$ & 0.193 & 1.420 \\ \hline
        aC--CNT(32,0) & 0.440 & 2.051 \\ \hline
        aC--CNT(32,0)$\dag$ & 0.220 & 2.051 \\ \hline\hline
    \end{tabular}
\captionof{table}{Lennard-Jones parameters used to model the interactions between aC and all entities in the system. The dagger symbol (\textdagger)  denotes terminal CNT beads.}
\label{tab:aCparams}
\end{minipage}
\end{figure*}

Amorphous carbon particles are also represented as single beads that interact with the other entities in the system (CNTs and other aC particles) via nonbonded Lennard-Jones potentials. Their interactions are parametrized with the same procedure used for CNTs, starting from the \textsc{airebo} potential~\cite{Stuart2000}. To this end, aC particles are first modeled as a collection of carbon atoms, randomly distributed within a spherical volume, with a minimum C--C distance corresponding to a double bond (0.134~nm) (see Fig.~\ref{fig:aCmodel}). The radius of the sphere is set to $0.6~\mathrm{nm}$, and the number of carbon atoms is adjusted to achieve a 2.5~g$\cdot$cm$^{-3}$ density. The particle is then placed at varying distances from another entity, and the total interaction energy is computed as the sum of all pairwise C--C \textsc{airebo} contributions. This calculation is repeated 300 times with different random internal configurations of the aC structure. The resulting interaction energy profiles are averaged and fitted to determine the two Lennard-Jones parameters $\varepsilon$ and $\sigma$ used in the classical force field (see Table~\ref{tab:aCparams}).

In this work, a single particle size and fixed density are adopted for all aC beads. Nonetheless, the same parametrization scheme can be extended to generate interaction parameters for a broad range of particle sizes and internal densities, enabling flexible modeling of diverse aC morphologies relevant to different experimental conditions and device architectures. 

\subsection{Creating CNT film models}\label{cnt_films}\label{method_CNT_films}

CNT film models were created by placing straight CNTs and aC particles at random positions and orientations within a large simulation box, corresponding to a low-density initial configuration. The lateral dimensions of the box $x = 200~\mathrm{nm}$ and $y=200~\mathrm{nm}$ remained fixed throughout the procedure, while the $z$ dimension was varied during compression to create a dense film. The system was confined between two parallel repulsive walls in the $z$ direction, whose interactions with the particles were described by a 12-6 Lennard-Jones potential. These walls represent the physical confinement imposed by the top and bottom electrodes. Periodic boundary conditions (PBCs) are applied only in the $xy$-plane.

To reach the target film density of either 0.3 or 0.6~g$\cdot$cm$^{-3}$, vertical downward forces are applied to all entities in the simulation box while gradually moving downward the upper wall (see SI for a description of the compression protocol). The dynamics of all particles is integrated in the canonical ($NVT$) ensemble, with the temperature controlled by a Nosé-Hoover thermostat set at $T=300~\mathrm{K}$. Once the desired volume is reached, the system is equilibrated at constant temperature until the total energy fluctuates around a steady negative value, after which a final energy minimization is performed to reach a local minimum. This procedure produces a single-layer CNT film contained within a $200 \times 200 \times 20 ~\mathrm{nm}^3$ simulation box.

Multilayer films are generated by first equilibrating four low-density layers which are then stacked together before the compression protocol is applied (see SI), thereby yielding films with the identical overall dimensions and density as their single-layer counterparts. As discussed in section~\ref{sec:results}, structures with higher density and CNT diameter are more difficult to compact. For these cases, several compression-decompression cycles with intermediate annealing stages at elevated temperature were required to reach the desired volume. 

\subsection{Structural metrics}\label{metrics}\label{method_metrics}

To analyze the structure of the obtained films, we used structural descriptors or metrics that offer a compact and interpretable representation of key morphological features and enable systematic exploration of how geometry and topology influence electron transport. 

The first metric quantifies the \emph{orientation} of the CNTs with respect to the $xy$-plane. Because CNTs may bend and buckle, the analysis is performed on each of their segments separately. From the unit direction vector $\mathbf{u}_i$ of a segment, the angle with the $xy$-plane is calculated from its $z$ component:
\begin{equation}
    \theta_i = \arcsin \left( \left| u_{i,z} \right| \right),
\end{equation}
where we take the absolute value to ensure that $\theta_i \in [0, \pi/2]$. The average $\theta$ and standard deviation $\sigma_{\theta}$ over all segments provide a measure of the global alignment with the confinement plane and of the orientational disorder, respectively.

The second metric is the \emph{buckling} factor $B$ simply defined as:
\begin{equation}
B = \frac{1}{N_{\textrm{CNT}}} \sum_j^{N_{\textrm{CNT}}}\frac{N_j^{\mathrm{buck}}}{N_j},
\end{equation}
where $N_j^{\mathrm{buck}}$ is the number of buckling points within the $j$-th CNT containing $N_j$ beads. $N_{\textrm{CNT}}$ is the total number of CNTs in the film. CNT beads are identified as buckling points when the segments they connect are misaligned by more than $\theta_B = 2.08^{\circ}$~\cite{zhigilei2005mesoscopic}.

\begin{figure*}[ht!]
\includegraphics[width=0.81\textwidth]{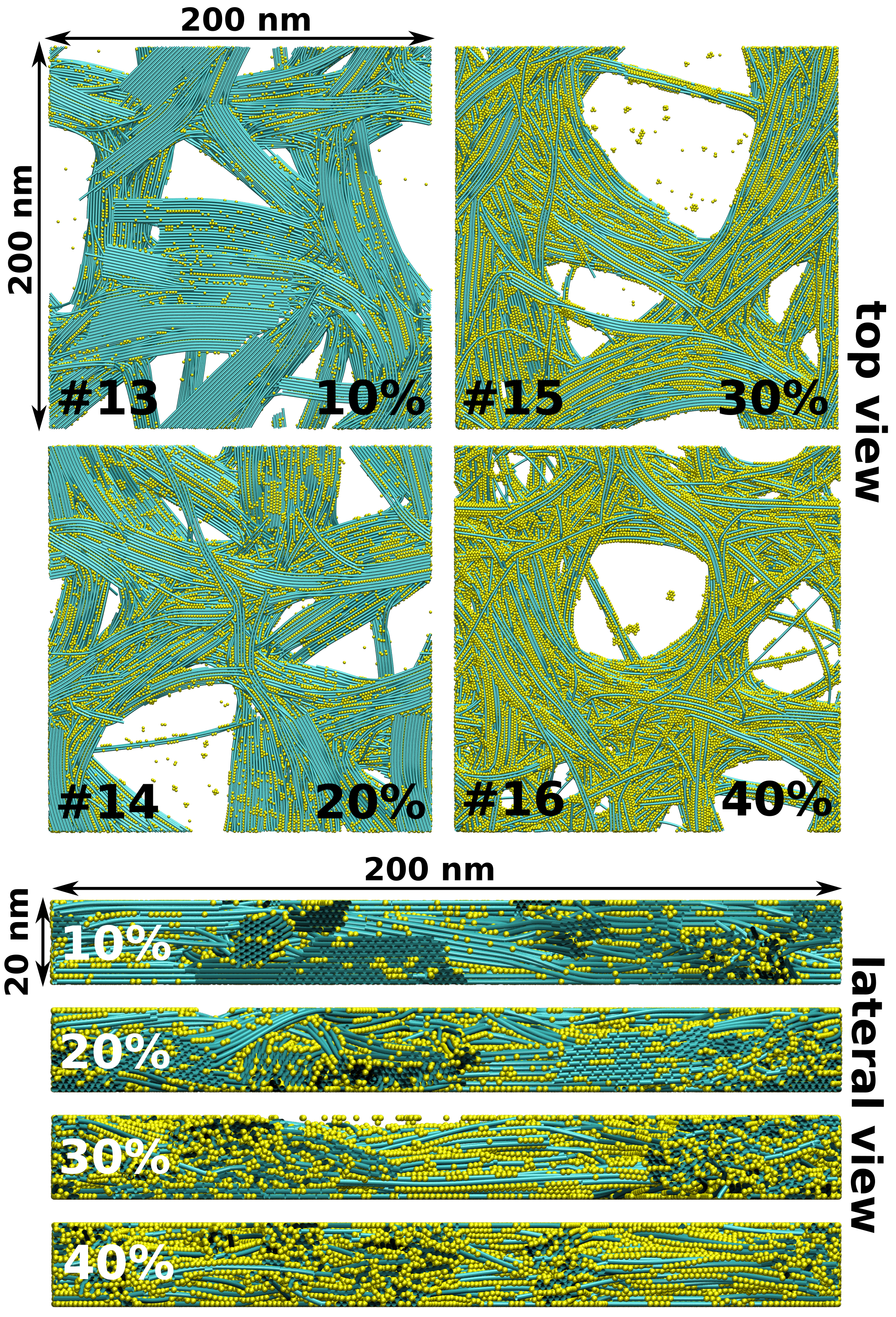}
\caption{Structures of single-layer films with a density of 0.6~g$\cdot$cm$^{-3}$ composed of (16,0), 100~nm-long CNTs containing between 10\% to 40\% amorphous carbon. CNT diameter is $d^{(16)} \simeq1.25~\mathrm{nm}$. CNT segments are shown in cyan and aC particles in yellow. (see Section~\ref{method_CNT_films} for details)}
\label{fig:structures_main}
\end{figure*}

The third metric is the \emph{curvature} factor $C$, which quantifies the average curvature of CNTs (excluding sharp kinks). First, we calculate the geometric curvature of the $j$-th CNT, at the position of its $i$-th bead:
 \begin{equation}
\kappa_j^i = \frac{||\mathbf{r}_i' \times \mathbf{r}_i''||}{||\mathbf{r}_i'||^3}
\end{equation}  
where the derivatives $\mathbf{r}_i'$ and $\mathbf{r}_i''$ are obtained using the finite difference method:
\begin{equation}
\mathbf{r}_i' = \frac{\mathbf{r}_{i+1} - \mathbf{r}_{i-1}}{2h} \quad ; \quad \mathbf{r}_i'' = \frac{\mathbf{r}_{i+1} - 2\mathbf{r}_{i} + \mathbf{r}_{i-1}}{h^2},
\end{equation}
where $h$ is the average between the two distances $r_{i-1,i}$ and $r_{i,i+1}$. The curvature for the first and the last beads takes the same values as the adjacent ones. The average curvature of the $j$-th CNT is then:
\begin{equation}
\kappa_j = \frac{1}{N_j - N_j^{\mathrm{buck}}} \sum_i^{\mathrm{non-buck}} \kappa_j^i
\end{equation}
where $N_j$ is the total number of beads in the CNT and $N_j^{\mathrm{buck}}$ the number of buckled beads. The average is only performed on beads that are not buckling points, since the curvature is discontinuous there and artificially large. The curvature factor $C$ for the entire film is then calculated from:
\begin{equation}
C = \frac{\sum_j \kappa_j}{N_{\textrm{CNT}}}.
\end{equation}

The fourth descriptor is a metric $\beta$ that quantifies \emph{bundling} in the CNT film. The method relies on a graph-based analysis of local CNT segment connectivity. First, all CNTs are divided into segments defined by two successive beads. Then each segment is associated with its midpoint position $\mathbf{r}_i$, a unit orientation vector $\mathbf{u}_i$, and the CNT index to which it belongs. A segment $i$ is considered connected to a segment $j$ if the following conditions are met: (\textit{i}) $i$ and $j$ belong to different CNTs, (\textit{ii}) the distance between midpoints is less than $1.5 \times d_0$, where $d_0$ is the CNT diameter (\textit{spatial proximity}), (\textit{iii}) the longitudinal projection $\Delta =  \left| (\mathbf{r}_j - \mathbf{r}_i) \cdot \mathbf{u}_i  \right|$ is below a threshold of $\sim0.9~\mathrm{nm}$ (the length of a segment is $\sim1.0~\mathrm{nm}$) to avoid false positives from connections to the extremities of the CNTs (\textit{axial overlap}), and (\textit{iv}) the angle between $\mathbf{u}_i$ and $\mathbf{u}_j$ is smaller than the buckling angle $\theta_B$. To prevent artificially inflated connectivity due to segment discretization, each segment is limited to at most six neighbors (the maximum number of hexagonal contacts in a bundle), and no multiple connections between the same pair of segments are allowed. Once the segment graph is built, its connected components are identified. Each component corresponds to a cluster of bundled CNT segments. The bundling metric $\beta$ is then defined as the average component size minus one (so that isolated segments yield zero bundling):
\begin{equation}
\beta = \frac{1}{N}\sum_{k=1}^{N_c} \left( n_k - 1 \right),
\end{equation}
where $N$ is the total number of segments, $N_c$ the number of connected components, and $n_k$ the size (number of segments) of the $k$-th component. This metric reflects the average number of connected neighbors per segment and captures local and global bundling in the film. We also report the average bundle size $\langle n \rangle = \frac{1}{N_c} \sum_k n_k$ as an additional indicator of the aggregate structure. In the limiting case where all CNT segments belong to a single connected component (\textit{i.e.}, one large bundle spanning the entire system) and assuming that PBCs also apply in the $z$-direction, the bundling metric would reach its maximum value $\beta = 1$. In the opposite case where no segment is connected to any other (\textit{i.e.}, all segments are isolated) one obtains $\beta = 0$. We also checked that small variations of these thresholds do not lead to discontinuous changes in the resulting metric.

Finally, to characterize the electrical connectivity of the network, we analyze the ensemble of current-carrying paths (see section~\ref{method_current}). Each path corresponds to a sequence of connected nodes (CNT beads) linking the two electrodes. From this set, we define two descriptors that quantify the ability of the network to conduct electricity. First, \emph{effective connectivity} $C_{\mathrm{eff}}$ is a weighted average reflecting the number and quality of conductive paths. Each path contributes a weight inversely proportional to the square of the number of internal junctions (excluding the two electrode contacts), thus penalizing long and tortuous paths:
\begin{equation}
C_{\mathrm{eff}} = \frac{1}{N_{\mathrm{paths}}} \sum_{\mathrm{paths}} \frac{1}{n_{\mathrm{junc}}^2},
\end{equation}
where $n_{\mathrm{junc}}$ is the number of internal junctions in the path. This descriptor favors short and direct conduction pathways. Second, \emph{the minimum number of junctions} $m$ corresponds to the smallest number of internal junctions observed across all conductive paths and provides a lower bound on the topological complexity required for charge transport through the network. These descriptors allow us to distinguish between highly interconnected films supporting efficient conduction and sparse networks, where the current must percolate through long paths. The last descriptor is simply the smallest number of connections $C_{\mathrm{el}}$ with the bottom or the top electrode.

\begin{figure*}[ht!]
\includegraphics[width=0.81\textwidth]{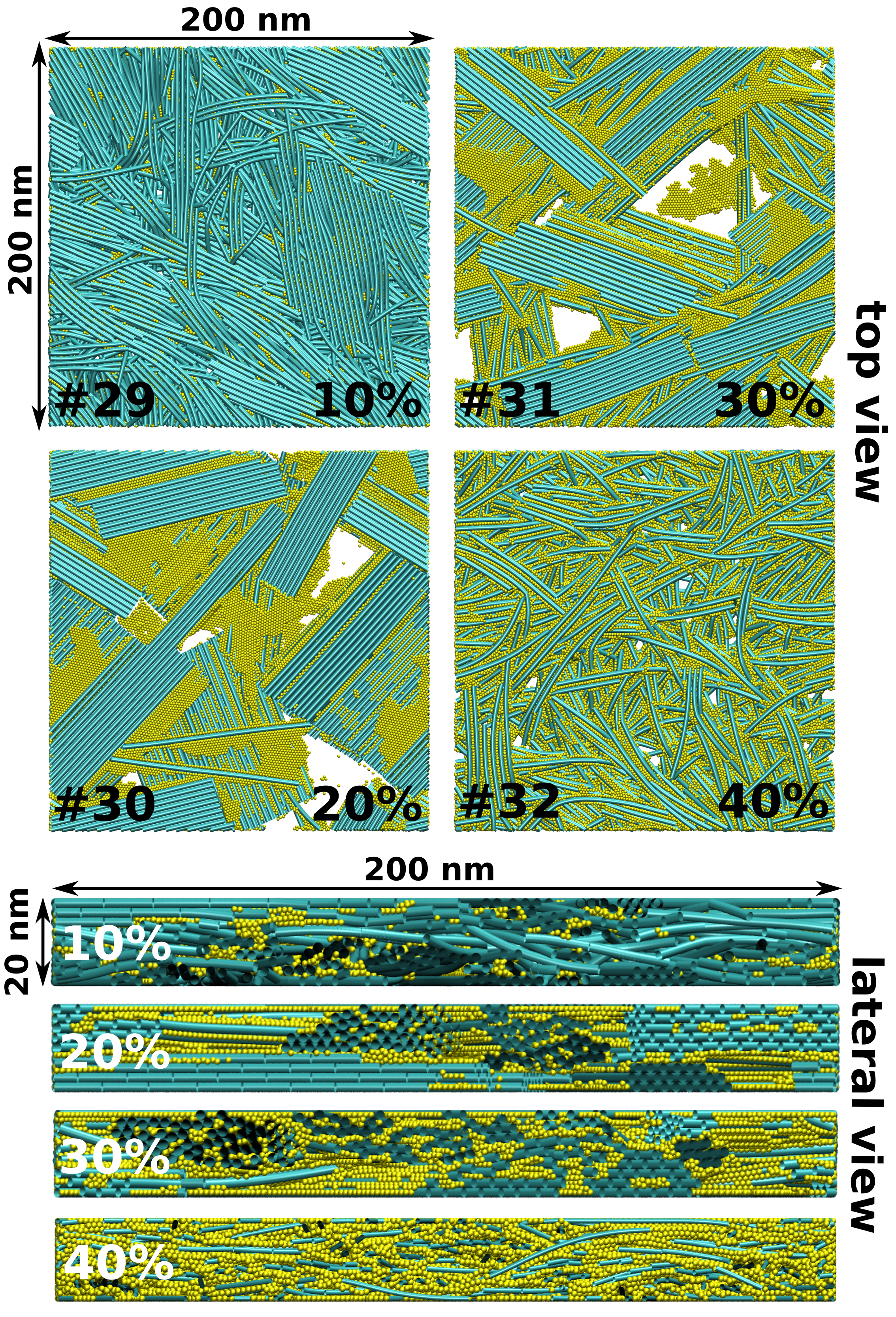}
\caption{Structures of single-layer films with a density of 0.6~g$\cdot$cm$^{-3}$ composed of (32,0), 100~nm-long CNTs containing between 10\% to 40\% amorphous carbon. CNT diameter is $d^{(32)} \simeq2.5~\mathrm{nm}$. CNT segments are shown in cyan and aC particles in yellow. (see Section~\ref{method_CNT_films} for details)}
\label{fig:structures_main2}
\end{figure*}

\subsection{Electrical current calculation}\label{method_current}

To compute the electrical current flowing through the carbon nanotube networks, we employ a nodal analysis approach, adapted from the work of Jin~\textit{et al.}~\cite{jin2018microstructural}. Each CNT bead is treated as a node, and electrical connections between nodes are encoded into a sparse conductance matrix $\mathbf{G}$. The overall structure is modeled as a resistor network where both intra-tube and intertube conductances are explicitly accounted for. We define a local intra-tube conductance $g_{\mathrm{intra}}$ between adjacent beads of the same nanotube, and an intertube conductance $g_{\mathrm{inter}}$ between connected segments belonging to different CNTs. 

Defect-free CNTs behave as one-dimensional ballistic conductors with a mean free path that exceeds $1~\mu\mathrm{m}$ for the bias regime considered here~\cite{park2004electron} and a conductivity of $2 G_0$ independently of the tube length~\cite{frank1998carbon,durrant2022atomistic}, where $G_0 \simeq 7.748\times 10^{-5}~\mathrm{S}$ is the quantum conductance. The intra-tube conductance is therefore given by:
\begin{equation}
g_{\mathrm{intra}} = 2 \left( N_{\mathrm{nodes}} - 1  \right)G_0,
\end{equation}
where $N_{\mathrm{nodes}}$ is the number of beads per CNT. 

Based on the model of Durrant \textit{et al.} combining density functional tight-binding (DFTB) theory and the non-equilibrium Green functions (NEGF) method~\cite{durrant2022atomistic}, we estimate the average conductance at the junctions between CNTs for both chiralities as:
\begin{eqnarray}
g_{\mathrm{inter}}^{(16,0)} &\simeq& 4.6083 \times 10^{-6}~\mathrm{S}  \nonumber \\
\quad \nonumber \\
g_{\mathrm{inter}}^{(32,0)} &\simeq& 5.7803 \times 10^{-7}~\mathrm{S}.
\end{eqnarray}
We note that tubes with larger diameters have lower junction conductance, an effect attributed to the reduced overlap between the delocalized conduction channels and the local contact area: as the tube circumference increases, the fraction of the electronic wave function contributing to the actual contact region decreases~\cite{durrant2022atomistic,buldum2001contact}. 

For the junctions between CNTs and electrodes, a high conductance of 0.05~S is assigned. CNT segments are considered connected when the shortest separation between the bead-bead lines is less than a cutoff radius of $r_{\mathrm{c}}^{(16,0)} \simeq 1.747$~nm~or $r_{\mathrm{c}}^{(32,0)} \simeq 3.005$~nm. These values correspond to twice the CNT radius plus the equilibrium van der Waals spacing in vacuum, augmented by a additional 0.15~nm. We verified that replacing this hard cutoff with an exponentially decaying distance-dependent conductance did not produce substantial differences in the total current. CNTs are also considered connected to the bottom or top electrode when their distance is less than $r_{\mathrm{c}}^{\mathrm{elec}} = 1.0$~nm. PBCs are applied in the plane of the film to ensure the correct treatment of finite-size effects, and special care is taken to correctly unwrap segment coordinates across the simulation box when identifying junctions.

Each junction is modeled by a contact conductance, which is incorporated into the matrix $\mathbf{G}$ according to Kirchhoff's current law. The external electrodes are represented as boundary nodes connected directly to a subset of CNTs. Once the global sparse matrix is assembled, we solve the following linear system:
\begin{equation}\label{linear}
    \mathbf{G} \cdot \mathbf{V} = \mathbf{s},
\end{equation}
where $\mathbf{V}$ is the electrostatic potential at each node and $\mathbf{s}$ encodes the boundary conditions imposed by the electrodes, with a bias of 0.5~V applied at the bottom electrode. The local electrical current flowing between two nodes $i$ and $j$ is given by:
\begin{equation}
    I_{ij} = G_{ij} \left( V_i - V_j \right).
\end{equation}
As the system is modeled as a purely resistive network of Ohmic CNTs obeying Kirchhoff's law, the total current transmitted through the film is conserved. It is therefore computed by summing all currents crossing the $xy$-plane at mid-height. 

We considered two cases: (\textit{i}) all CNTs are metallic, and (\textit{ii}) since many CNT chiralities are semiconducting, only one-third of the tubes are randomly selected as being metallic, while the remaining semiconducting ones are assumed to be non-conductive under the 0.5~V bias applied here. While specific chiralities are used to define CNT geometries in the film construction, the subsequent electrical transport calculations are independent of chirality and depend only on the nanotube diameter. In the present model, the electrical conductance of each connection is treated as Ohmic and independent of the carrier statistics, so that the fully metallic and partially metallic networks follow the same linear transport scheme. In reality, when only one-third of the CNTs are metallic, the system would likely operate in a mixed regime where quantum conduction along metallic paths coexists with thermally activated hopping across semiconducting barriers~\cite{yanagi2010transport,jiang2018ultrahigh}. Such effects are not explicitly captured here, but are expected to primarily rescale the overall conductance without altering the qualitative trends observed between structural descriptors and transport efficiency. 

After solving the linear system given by Eq.~(\ref{linear}), we extract the set of bonds that carry a significant current and reconstruct the connected subgraph linking the source and drain electrodes to identify the available dominant conduction pathways within the network. We treat aC particles as electrically inactive inclusions, affecting the network morphology but not contributing directly to the total film conductivity, under the assumption that their conductivity is much lower than that of CNTs; their inclusion as conductive elements can nevertheless be straightforwardly implemented within the same framework in future work.

\section{Results of Calculations}\label{sec:results}
\subsection{Film structure}

Structures containing amorphous carbon are shown in Figs.~\ref{fig:structures_main} and~\ref{fig:structures_main2} while all other configurations are provided in the SI (Figs.~S3 and~S4). The complete set of calculated descriptors and currents is also presented in the SI (Table~S1). 

Due to the inherent variability in the initial random configurations and the necessity of applying a slightly different compression protocol for structures \#30 and \#31 (combining high-density, intermediate amorphous carbon content and large-diameter tubes), the quantitative and physical interpretation of the descriptor correlations remains exploratory. A statistically more robust analysis would require multiple realizations per condition and a more uniform compression protocol across all structures. Nevertheless, the consistent qualitative trends observed across the dataset, in particular for buckling and bundling, indicate that the selected set of descriptors captures relevant topological and geometrical features of the CNT networks and provides a useful basis for future, more systematic studies. Structures \#30 and \#31 developed a different morphology compared to other aC-containing films, illustrating that under additional compression cycles the protocol can access distinct morphologies with enhanced bundling and aC aggregation.

Low-density films composed of short CNTs (15 or 40 nm in length; \#1$\rightarrow$\#4 and \#17$\rightarrow$\#20) tend to exhibit large voids and pronounced bundle formation, reflected in high values of the bundling factor $\beta$ and average bundle size $\langle n \rangle$, while showing no detectable buckling ($B=0$). According to the stability criterion proposed by Volkov \textit{et al.}~\cite{volkov2010structural}, which identifies stable networks as those that avoid fragmentation into isolated and weakly connected bundles, these systems would be considered unstable. These configurations consistently show poor effective connectivity ($C_{\mathrm{eff}}$) and low electrical conductivity, particularly for high value of $\langle n \rangle$. For low-density films with 100-nm long tubes we observe a reduction in bundling and a concurrent increase in bending, curvature, and current for the (16,0) chirality. These trends remain nearly unchanged for the large-diameter (32,0) tubes, showing that thinner CNTs are more prone to mechanical deformation under compression, which in turn improves intertube contacts and electrical conduction.

In contrast, high-density structures exhibit significantly fewer voids, especially in the four-layer configurations and for the (32,0) chirality. This trend is not surprising. On the one hand, (32,0) films at 0.6~g$\cdot$cm$^{-3}$ are already close to their close-packed density, whereas the corresponding value for (16,0) films is approximately 1.5 times higher, leaving more space for void formation. However, since the position of the voids within each layer is random, the voids do not align when the layers are stacked together. 

We note that without confinement in the $z$-direction, several of these high-density films would expand and adopt a lower-density structure. If film stability were judged on the basis of their ability to maintain their morphology in vacuum, some high-density films would likely be considered unstable. However, in the capacitor geometry considered here, the lateral confinement is intrinsic and the film density remains constant throughout. For these dense morphologies, layering systematically reduces electrical conductivity, whereas this trend is not observed for their low-density counterparts. The role of amorphous carbon appears to be more versatile, with its influence sensitively depending on the combined effects of chirality, mass fraction, and compression protocol. These observations highlight the complex interplay between structural parameters, such as chirality, tube length, density, and amorphous carbon content, which collectively dictate the final morphology and transport performance of the network.

\begin{figure*}[t!]
\centering
\begin{subfigure}{.5\textwidth}
  \centering
  \includegraphics[width=\linewidth]{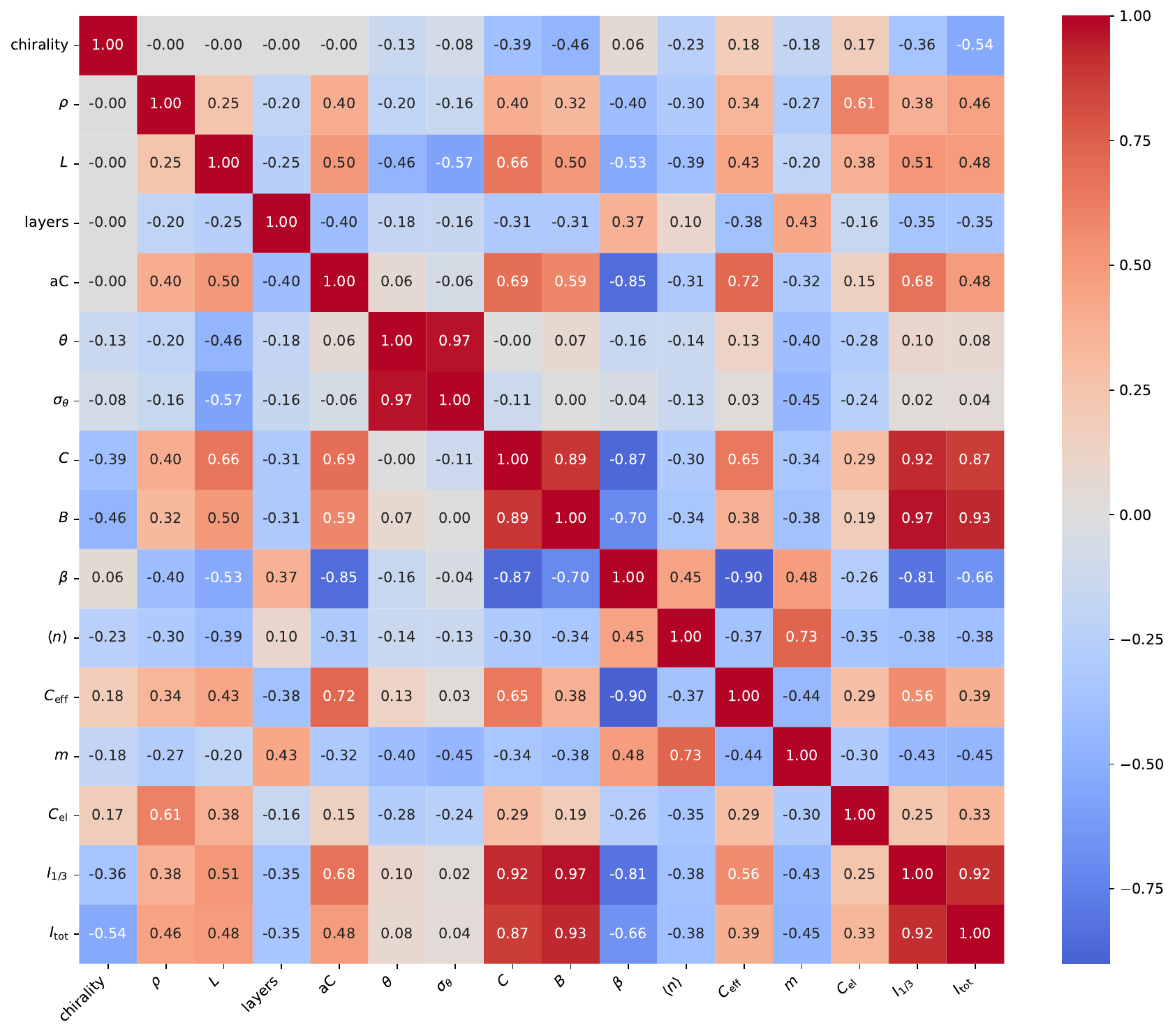}
  \caption{Pearson correlation matrix}
  \label{fig:pearson}
\end{subfigure}%
\begin{subfigure}{.5\textwidth}
  \centering
  \includegraphics[width=\linewidth]{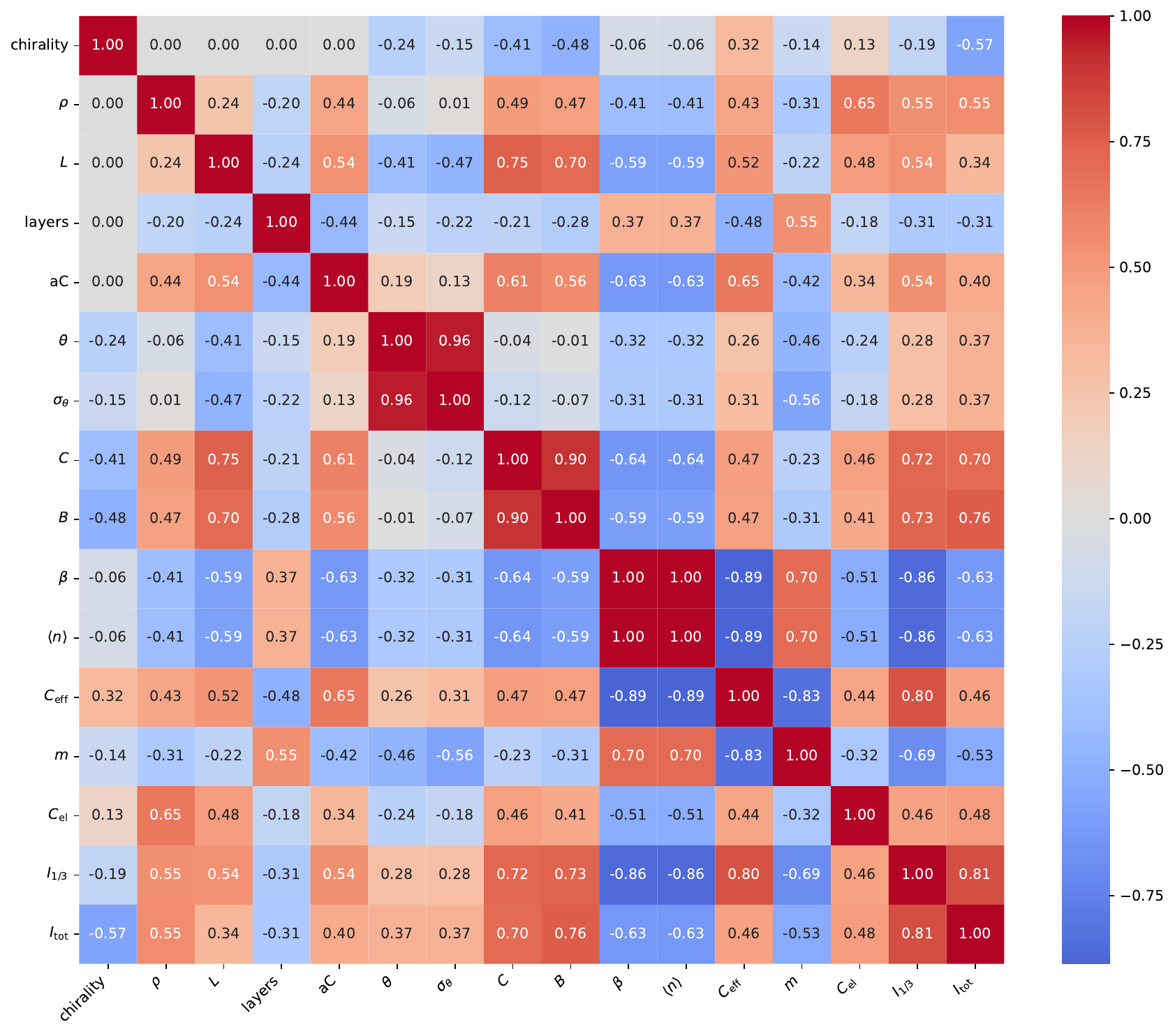}
  \caption{Spearman correlation matrix}
  \label{fig:spearman}
\end{subfigure}
\caption{Correlation matrices between the different descriptors and the currents.}
\label{fig:correlation_matrices}
\end{figure*}

\subsection{Correlation between structure and electrical transport}

To quantitatively assess the relationships between structural descriptors and transport properties, we first computed the Pearson correlation matrix~\cite{benesty2009noise} across all available parameters, as shown in Fig.~\ref{fig:pearson}. Several strong correlations emerge, providing insight into how geometry and topology affect the electrical performance of CNT networks. Local geometric distortions, captured by the curvature factor $C$ and buckling factor $B$, are among the most influential descriptors. These two quantities are strongly correlated ($r=0.89$), as both reflect geometric distortions at the nanotube segment level. In particular, both $C$ and $B$ show a strong positive correlation with the total current $I_{\mathrm{tot}}$ ($r=0.87$ and $r=0.93$, respectively) and even more with the current $I_{1/3}$ when only one-third of the tubes are considered metallic ($r=0.92$ and $r=0.97$). These results suggest that local mechanical distortions facilitate charge transport, probably by enhancing intertube overlap and contact and/or by reducing the number of junctions along the main conductive paths.

In contrast, the bundling factor $\beta$ shows a strong negative correlation with $I_{\mathrm{tot}}$ ($r = -0.66$) and with $I_{1/3}$ ($r=-0.81$), indicating that highly bundled networks tend to conduct less efficiently. Similar trends have been widely reported in low-density transparent CNT films, where increasing the size of the bundles reduces the number of effective intertube junctions and limits charge percolation~\cite{hecht2006conductivity, lyons2008relationship, simien2008influence, nirmalraj2009electrical,jiang2018ultrahigh,shin2009conductivity}. The geometric arguments made in these studies are consistent with the behavior we observe in our high-density films.

Effective connectivity $C_{\mathrm{eff}}$ exhibits a moderate positive correlation with both currents ($r=0.56$ with $I_{1/3}$ and $r=0.39$ with $I_{\mathrm{tot}}$), indicating that global network connectivity contributes to transport, though to a lesser extent than local geometric factors such as curvature and buckling, which show stronger correlations. This highlights the combined influence of local and global structural features on charge conduction. Interestingly, the minimum number of connections to the electrodes shows only a weak correlation ($r=0.25$ and $r=0.33$, respectively), which does not rule out a more significant role of the electrodes under cycling or nonequilibrium conditions.

Global structural parameters also influence connectivity. The length of the tube is positively correlated with curvature ($r=0.66$), buckling ($r=0.50$), $C_{\mathrm{eff}}$ ($r=0.43$), and both currents ($r=0.48$ for $I_{\mathrm{tot}}$ and $r=0.51$ for $I_{1/3}$). Longer tubes are more likely to establish multiple contacts and undergo elastic deformation, thereby improving network connectivity. Film density correlates moderately with $C$, $B$, and $C_{\mathrm{eff}}$ ($r\sim0.3-0.4$), as well as with total current ($r=0.46$), consistent with the expected increase in intertube interactions in denser networks. In contrast, the number of layers is anticorrelated with connectivity and transport ($r=-0.35$ with currents, $r=-0.38$ with $C_{\mathrm{eff}}$), likely reflecting the tendency of multilayer films to introduce additional junctions along percolation paths. Finally, the tube orientation angle $\theta$ and its dispersion $\sigma_{\mathrm{\theta}}$ are highly correlated with each other ($r=0.97$), but show negligible correlation with transport properties or connectivity descriptors, indicating that the macroscopic angular disorder does not significantly affect global transport in the present dataset. However, the effects of local alignment or anisotropic percolation pathways may still play a role beyond the scope of the linear correlation analysis presented here.

The role of amorphous carbon is more complex and highly structure-dependent. Although it correlates positively with curvature ($r=0.69$) and buckling ($r=0.59$), its correlation with transport is less clear. No monotonic trend is observed between the aC content and the current. This likely reflects the inhomogeneous spatial distribution of aC and the non-uniform generation protocols. In particular, structures with chirality (32,0) and 20-30\% aC content (\#30 and \#31) required additional pressure-compression cycles, leading to the formation of large aC clusters and modified bundle morphologies. Such features likely affect transport independently of the nominal aC percentage. For (16,0) structures (\#13-\#16), aC consistently aligns along CNTs, while (32,0) structures exhibit more diverse behaviors: junction clustering at 10\% aC (\#29), plaque formation at 20-30\% (\#30-\#31), and tube alignment at 40\% (\#32). These variabilities highlight the sensitivity of aC morphology to both diameter and processing history. Consequently, conclusions about the role of aC should be drawn with caution. A more rigorous assessment would require controlled structure generation and multiple realizations per configuration to statistically capture the influence of aC on film morphology and transport. These observations also suggest that it should be possible to tune the morphology of aC-containing films through careful control of the construction protocol, enabling modeling of different experimental scenarios depending on the targeted application or physical conditions. Finally, we note that the aC content is generally expected to affect the conductivity of the film not only through the morphological changes discussed here, but also through additional carrier scattering~\cite{geng2007dependence}, which would tend to reduce transport efficiency, or through the opening of thermally activated hopping pathways, which could have a competing effect.

Looking ahead, explicitly incorporating hopping conduction through aC phases offers a promising route to more accurately describe and ultimately engineer charge transport in CNT films. Beyond its role as a passive matrix, aC may actively mediate intertube connectivity via thermally activated hopping pathways, particularly in regimes where direct tube-tube
contact or tunnelling is limited. Systematic exploration of aC concentration and cluster morphology ranging from isolated nanoscale domains to percolating networks could reveal non-trivial transitions between insulating, hopping-dominated, and quasi-metallic transport regimes. Such control is likely to impact not only electrical conductivity but also film uniformity, mechanical integrity, and environmental stability. Ultimately, integrating these effects into multiscale models and experimental design frameworks could enable the rational optimization of CNT films for applications in flexible electronics, interconnects, and sensing technologies.

\begin{figure}[ht!]
\includegraphics[width=0.8\columnwidth]{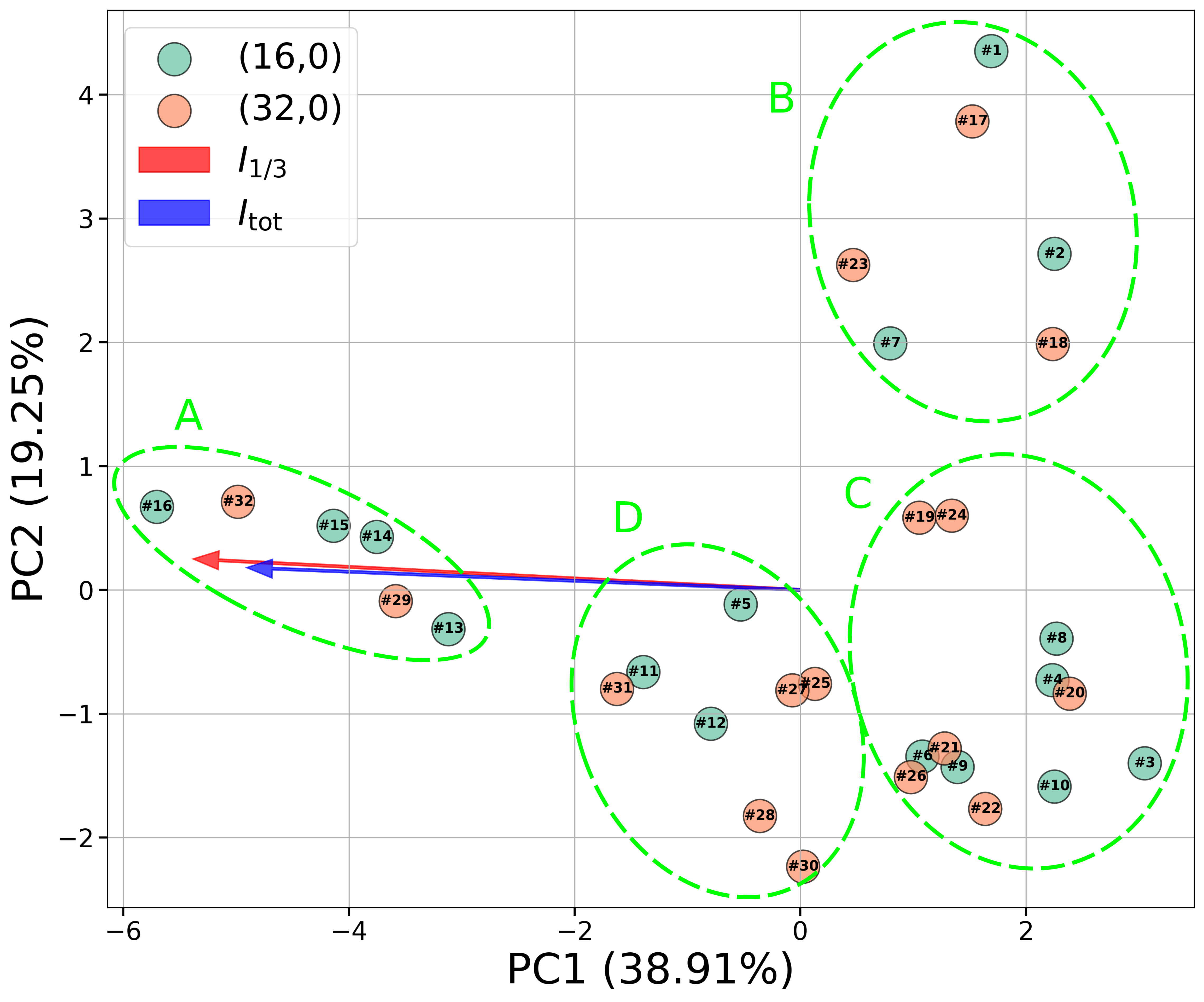}
\caption{Principal Component Analysis of structural descriptors. Each point represents one structure. The projection is shown in the plane of the first two principal components, PC1 and PC2. Arrows indicate the directions of correlation with transport properties. Cluster A includes only films with 100 nm-long tubes containing aC. Cluster B contains only films with 15 nm-long tubes. Cluster C mainly consists of films with either 15 or 40 nm-long tubes, along with a few low-density films with 100 nm-long tubes. Cluster D consists mostly of films with 100 nm-long tubes without aC, except for structures \#30 and \#31.}
\label{fig:pca}
\end{figure}

\subsection{Trend analysis}

To test the robustness of the trends identified with the Pearson correlation matrix, we also calculated the Spearman rank correlation matrix~\cite{hauke2011comparison}, as shown in Fig.~\ref{fig:spearman}, which captures monotonic and possibly nonlinear relationships between variables. Overall, the Spearman analysis confirms most of the trends previously observed, particularly the strong associations between curvature, buckling, and transport properties. Some correlations involving the amorphous carbon content were slightly reduced, suggesting that its influence on transport is not strictly linear. In contrast, the correlations between angular descriptors ($\theta$ and $\sigma_{\theta}$) and current values were more pronounced by the Spearman analysis, hinting at potential nonlinear effects of tube orientation that are not well captured by Pearson correlation. This observation suggests that local alignment may influence transport pathways in a more subtle, nonlinear manner. The overall consistency between the two analyses reinforces the relevance of the identified descriptors, while also emphasizing the importance of considering nonlinear interactions in future modeling efforts. While pairwise Pearson and Spearman correlation matrices provide valuable insight into individual descriptor–transport relationships, they do not capture the collective influence of multiple correlated descriptors. To address this, we next employ principal component analysis (PCA)~\cite{wold1987principal,abdi2010principal, jolliffe2011principal,greenacre2022principal}.

 Given the limited number of simulated structures, we performed a PCA analysis as an exploratory tool on the entire set of normalized structural descriptors, including geometrical and topological quantities. Electrical currents $I_{\mathrm{tot}}$ and $I_{1/3}$ were not included as input variables. However, their subsequent projection into the PCA space reveals how they correlate with the underlying structural patterns. The results for the most significant principal components are shown in Table~S2. This dimensionality reduction reveals dominant patterns of variation across the dataset and facilitates the interpretation of the underlying structure-property relationships. The first principal component (PC1) alone captures 38.9\% of the total variance, while PC2 accounts for an additional 19.3\%. Together, the first three components explain more than 71\% of the variance. 

Analyzing component loadings, PC1 is mainly associated with curvature, buckling, effective connectivity, and amorphous carbon content, and to a lesser extent with CNT length, density, and the number of connections to electrodes; all exhibiting negative coefficients. In contrast, features such as the bundling factor, mean bundle size, number of layers, and minimum number of connections to electrodes exhibit positive contributions. This implies that structures with large curvature and buckling, high connectivity, and minimal bundling, \textit{i.e.} those with strongly negative PC1 scores, are also those showing enhanced transport performance. Negative loadings of transport-enhancing descriptors simply reflect the arbitrary orientation of PC1 in the eigen-decomposition and do not imply an inverse physical relationship (see SI). This trend aligns with our earlier correlation analysis and confirms the central role of morphological order and network connectivity in enabling efficient conduction. PC2 is largely dominated by angular disorder and, to a lesser extent, by CNT length and the number of layers. These orientation descriptors exhibit limited correlation with transport, and their prominence along PC2 reflects a distinct source of structural variation unrelated to conductivity. Higher-order components (PC3 and beyond) capture more specific effects, such as chirality, bundling size, and layer count.

In the PCA projection along the PC1-PC2 plane, shown in Fig.~\ref{fig:pca}, individual structures can be grouped into clusters, in particular according to the length of tubes, with tightly grouped configurations observed for aC-containing structures (cluster A). Apart from structures \#30 and \#31, networks containing aC are the most conductive. Transport vectors ($I_\mathrm{tot}$, $I_{1/3}$), when added as supplementary arrows in the correlation plot, point strongly along the negative PC1 direction, confirming that well-connected, low-bundling structures with high curvature and buckling consistently produce higher current output. In contrast, structures with weak connectivity and strong bundling, characterized by positive PC1 scores, fall into the low-performance regime. These findings highlight the value of PCA not only as a tool for reducing dimensionality but also as a means to identify collective structural signatures that govern macroscopic transport.

\section{Discussion}\label{sec:Discussion}

We have presented a mesoscopic modeling framework for simulating high-density CNT networks with different structural morphologies and amorphous carbon content and analyzed their structural characteristics and electrical transport properties. A diverse set of 32 CNT films was constructed, systematically varying chirality, tube length, network density, and aC content. The resulting structures were characterized by a set of geometrical and topological descriptors, and electrical current through the films was calculated with a nodal analysis framework. Additionally, we investigated the role of layering by comparing single- and multilayer films and found that its impact on connectivity and transport depends significantly on the structural context, including chirality and tube length. To account for amorphous carbon, we introduced a simple yet effective modeling strategy enabling its integration into CNT films. Our analysis reveals that the aC content can significantly alter the local morphology and thereby influence electrical transport, although in a nonmonotonic and configuration-dependent manner. However, the presented modeling framework relies on a set of simplifications that define the range of applicability of the results reported here.

First, the simulated networks are composed exclusively of single-walled carbon nanotubes (SWCNTs), as the current coarse-grained model does not include an explicit representation of multiple concentric CNT walls or their associated mechanical response. As a result, the quantitative magnitude of curvature- and buckling-related effects reported in this work reflects the mechanical compliance of SWCNT-based networks. While the underlying mesoscale framework remains applicable in principle, extending the present analysis to multiwalled CNT (MWCNT) networks would require the explicit inclusion of MWCNT-specific mechanical properties, which may modify the relative weight of deformation-related descriptors without altering their qualitative role.

Secondly, amorphous carbon (aC) particles are treated here as electrically inactive inclusions that influence the network morphology but do not contribute directly to electrical transport. This modeling choice allows us to isolate the impact of aC on CNT packing, bundling, and junction statistics. In the generated structures, aC particles are predominantly attached to CNT sidewalls rather than occupying intertube junction regions, so that the dominant current pathways remain CNT-mediated. Neglecting the conductive contribution of aC may therefore affect the quantitative distribution of current and the detailed topology of percolating paths in systems where aC participates actively in transport. However, given their size and spatial arrangement, aC-mediated pathways would typically involve a larger number of junctions, making them less efficient than CNT-dominated routes within the present structural regime. This modeling choice is therefore physically justified for the configurations investigated here.

Thirdly, the structural descriptors introduced in this study are analyzed within a static framework and are used to characterize baseline structure-transport relationships. For the applications that involve dynamic, field-driven structural evolution, the descriptors identified here provide a natural basis for future extensions. For instance, field-induced filament formation or junction reconfiguration in RRAM would be expected to manifest as changes in effective connectivity $C_{\mathrm{eff}}$, in the minimum number of junctions along dominant current paths, or in the bundling factor $\beta$. In this context, the present results identify which descriptors are most relevant for transport and therefore which quantities are likely to be most sensitive to electrically driven structural evolution, without explicitly modeling switching dynamics.

Finally, the quantitative trends reported in this work pertain to a specific range of CNT lengths, network densities, and aC contents motivated by experimentally fabricated dense CNT films. In networks composed of substantially longer CNTs, increased aspect ratios would enhance bridging capabilities and increase the number of junctions per tube, potentially modifying the relative importance of local descriptors such as bundling. Extending the present analysis to such regimes would require additional global topological descriptors such as entanglement and tortuosity which are expected to become increasingly relevant. In particular, while increased entanglement or tortuosity may counterbalance the negative impact of bundling by improving global connectivity, the local effect of bundling in increasing the number of effective intertube junctions along a given pathway is expected to remain qualitatively unchanged. We believe that the underlying mesoscale modeling approach and the associated descriptor-based analysis presented here are general and can be readily applied to a broader class of CNT-based networks and composite architectures.

\section{Conclusions}\label{sec:conclusion}

Despite the inherent variability in initial configurations and compression protocols, our descriptor-based analysis reveals consistent trends linking morphological features to electrical transport. Curvature and buckling factors, which capture local mechanical distortions, show strong positive correlations with current, suggesting that deformation improves the contact between tubes and the effective conduction pathways. In contrast, bundling suppresses the current, most likely due to a reduction in overall connectivity across the network. Global descriptors such as CNT length and network density also play significant roles, reinforcing the idea that both local geometry and global topology jointly determine charge transport. Both Pearson and Spearman correlation matrices confirm these trends, highlighting strong monotonic and, in some cases, nonlinear dependencies between structural descriptors and electrical transport. The principal component analysis further consolidates these findings, with the first component defining a structural axis characterized by high curvature, strong buckling, low bundling, and high connectivity. This reduction in dimensionality highlights that efficient transport is governed by a concerted combination of factors rather than by any single descriptor.

In summary, our results demonstrate that high-density CNT films can be rationally designed to achieve specific target morphologies by tuning the amorphous carbon content and the compression protocol, with amorphous carbon incorporated as spherical domains of controllable size and density. We also demonstrate that a limited set of structural descriptors can effectively capture and explain the electrical behavior of such films.  This study lays the foundation for a more systematic exploration of structure–transport relationships in CNT-based networks and composite films and highlights the potential of mesoscale modeling as a predictive tool to guide the design of complex nanostructured materials.

\section*{Supporting Information} 

Experimental CNT length and diameter distributions in RRAM films; additional CNT film structures; compression protocol details; complete table of structural descriptors and calculated currents; details of Pearson, Spearman, and PCA analyses.

\section*{Acknowledgments}

We acknowledge the computational resources provided by the Aalto Science-IT project, and the use of the UCL Kathleen High Performance Computing Facility (Kathleen@UCL), and associated support services, in the completion of this work. The authors are grateful to P. A. Kloza and J. A. Elliott for valuable discussions and to A. N. Volkov and L. V. Zhigilei for introduction into CNT force fields and useful comments.

\bibliography{cnt_bib}

\clearpage
\newpage
\section*{Graphical abstract}

\begin{figure*}[ht]
\includegraphics[width=\textwidth]{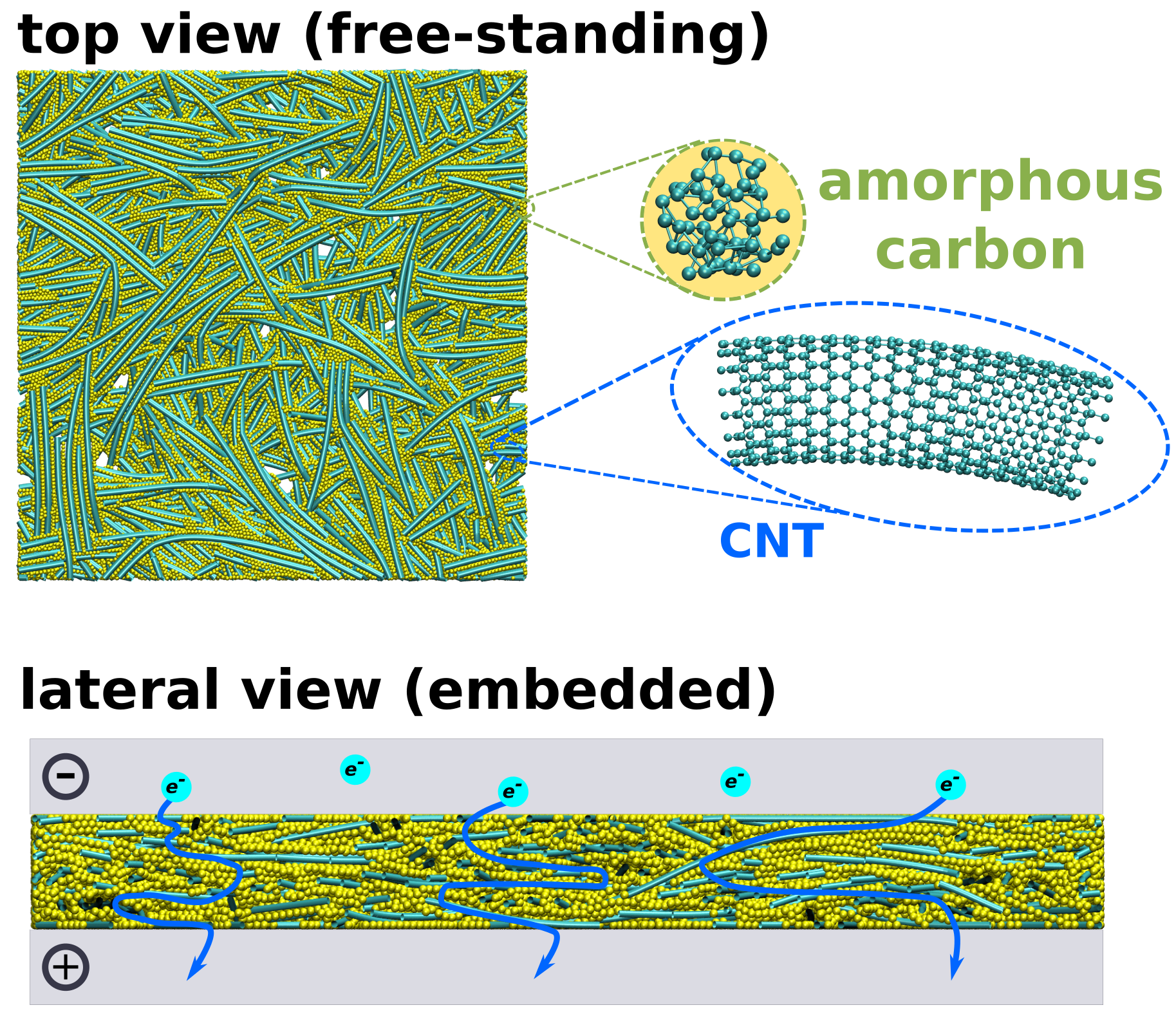}
\end{figure*}

\end{document}